\documentclass[aps, twocolumn, superscriptaddress, showpacs, nofootinbib, prd,longbibliography, citecolor=blue,colorlinks=true]{revtex4-2}
\usepackage[utf8]{inputenc}
\usepackage[T1]{fontenc}
\usepackage{ae,aecompl} 
\usepackage{graphicx}
\graphicspath{{figures/}} 
\usepackage{amsmath}
\usepackage{color}
\usepackage{xcolor}
\usepackage{amssymb}
\usepackage{orcidlink}
\usepackage{latexsym}
\usepackage{wasysym}
\usepackage{psfrag}
\usepackage{ifthen}
\usepackage{hyperref}
\usepackage{longtable}
\usepackage{float}
\usepackage[utf8]{inputenc}
\usepackage{lineno}
\usepackage{units}
\usepackage{multirow}
\usepackage{ulem}
\def\apjl{Astrophys.~J.~Lett}
\def\mnras{Mon. Not. R. Astron. Soc.}

\begin{abstract}
We present an all-sky search for long-duration gravitational waves (GWs) from the first part of the LIGO-Virgo-KAGRA fourth observing run (O4), called O4a and comprising data taken between 24 May 2023 and 16 January 2024. 
The GW signals targeted by this search are the so-called ``long-duration'' ($\gtrsim 1$\,s) transients expected from a variety of astrophysical processes, including non-axisymmetric deformations in magnetars or eccentric binary  coalescences. We make minimal assumptions on the emitted GW waveforms in terms of morphologies and durations. Overall, our search targets signals with durations $\sim 1$--$1000$\,s and frequency content in the range $16$--$2048$\,Hz. In the absence of significant detections, we report the sensitivity limits of our search in terms of root-sum-square signal amplitude ($h_\textrm{rss}$) of reference waveforms. These limits improve upon the results from the third LIGO-Virgo-KAGRA observing run (O3) by about 30\% on average.  
Moreover, this analysis demonstrates substantial progress in our ability to search for long-duration GW signals
owing to enhancements in pipeline detection efficiencies. As detector sensitivities continue to advance and observational runs grow longer, unmodeled long-duration searches will increasingly be able to explore a range of compelling astrophysical scenarios involving neutron stars and black holes.
\end{abstract}

\begin{document}
\title{All-sky search for long-duration gravitational-wave
transients in the first part of the fourth LIGO-Virgo-KAGRA
Observing run}


\author{A.~G.~Abac\,\orcidlink{0000-0003-4786-2698}}
\affiliation{Max Planck Institute for Gravitational Physics (Albert Einstein Institute), D-14476 Potsdam, Germany}
\author{I.~Abouelfettouh}
\affiliation{LIGO Hanford Observatory, Richland, WA 99352, USA}
\author{F.~Acernese}
\affiliation{Dipartimento di Farmacia, Universit\`a di Salerno, I-84084 Fisciano, Salerno, Italy}
\affiliation{INFN, Sezione di Napoli, I-80126 Napoli, Italy}
\author{K.~Ackley\,\orcidlink{0000-0002-8648-0767}}
\affiliation{University of Warwick, Coventry CV4 7AL, United Kingdom}
\author{C.~Adamcewicz\,\orcidlink{0000-0001-5525-6255}}
\affiliation{OzGrav, School of Physics \& Astronomy, Monash University, Clayton 3800, Victoria, Australia}
\author{S.~Adhicary\,\orcidlink{0009-0004-2101-5428}}
\affiliation{The Pennsylvania State University, University Park, PA 16802, USA}
\author{D.~Adhikari}
\affiliation{Max Planck Institute for Gravitational Physics (Albert Einstein Institute), D-30167 Hannover, Germany}
\affiliation{Leibniz Universit\"{a}t Hannover, D-30167 Hannover, Germany}
\author{N.~Adhikari\,\orcidlink{0000-0002-4559-8427}}
\affiliation{University of Wisconsin-Milwaukee, Milwaukee, WI 53201, USA}
\author{R.~X.~Adhikari\,\orcidlink{0000-0002-5731-5076}}
\affiliation{LIGO Laboratory, California Institute of Technology, Pasadena, CA 91125, USA}
\author{V.~K.~Adkins}
\affiliation{Louisiana State University, Baton Rouge, LA 70803, USA}
% ...

\collaboration{The LIGO Scientific Collaboration, the Virgo Collaboration, and the KAGRA Collaboration}



\date[\relax]{compiled \today}
\maketitle

\section{Introduction}
The direct detection of gravitational waves (GWs) from a pair of black holes (BHs) during the first observing run (O1)  \cite{2016PhRvL.116f1102A} of the Laser Interferometer Gravitational-wave Observatory (LIGO) \cite{2015CQGra..32g4001L}   marked a major milestone in GW astrophysics. The detection of GWs from the binary neutron star (NS) merger GW170817 during the second LIGO and Virgo \cite{2015CQGra..32b4001A}  observing run (O2) \cite{2017PhRvL.119p1101A} initiated the era of multi-messenger astrophysics, via the identification of an electromagnetic counterpart in practically all bands of the spectrum by a suite of electromagnetic observatories from the ground and space \cite{2017ApJ...848L..13A,2017ApJ...848L..12A}.

During the third observing run (O3) of Advanced LIGO, Virgo, and KAGRA \cite{2019NatAs...3...35K,2021PTEP.2021eA102A},  several other interesting compact binary systems were detected. These include a likely NS-NS merger with total mass significantly larger than that of known galactic binary NS systems \cite{2020ApJ...892L...3A}; BH-NS candidates \cite{2021ApJ...915L...5A}; pairs of compact objects with one of the two belonging to the so-called lower mass gap (the dividing line between the heaviest NSs and the lightest BHs) \cite{2020ApJ...896L..44A} and a BH-BH coalescence whose compact remnant falls in the intermediate BH mass range \cite{2020PhRvL.125j1102A}. 
In total, about $90$ transient GW signals have been confidently detected in the O1-O3 runs, all of them confidently associated with compact binary coalescences (CBC) \cite{GWTC-1, GWTC-2, GWTC-3}.
The O4 run is the longest LIGO-Virgo-KAGRA observing run to date. Similarly to the O1-O3 runs, the first eight months of O4 (called O4a) have brought new CBC detections, including the notable GW230529---a merger of a NS with a lower mass-gap BH \cite{2024ApJ...970L..34A}. 

While all of the above mentioned O1--O4 detections are exciting, we are yet to probe the large variety of astrophysical scenarios that predict GWs from systems other than CBCs. This motivates continued searches for other classes of GW signals (e.g., \citep{2018CQGra..35f5009A,2019PhRvD..99j4033A, 2021PhRvD.104j2001A,2022PhRvD.106j2008A,2023ApJ...955..155A,2024ApJ...966..137A,2025ApJ...985..183A}). 
Here, we present a search for unmodeled, long-lived (\unit[$\sim 1$---$1000$]{s}) GW transients in O4a data, updating results from the first three runs \cite{2018CQGra..35f5009A,2019PhRvD..99j4033A, 2021PhRvD.104j2001A}. 

This analysis includes the first application of the XGBoost \cite{XGBoost} post-processing classifier of coherent WaveBurst (cWB) to long-duration GW searches (Section \ref{Sec:method}), and results from the first PySTAMPAS \citep{2021PhRvD.104j2005M} all-sky search for long-duration GWs (Section \ref{Sec:method}). 
The use of diverse methods in the search for long-duration GW signals is motivated by the wide range of potential signal morphologies---including variations in duration, temporal evolution, and frequency content---expected from different astrophysical scenarios. Many of these scenarios remain unexplored and constraining them via GW observations holds the promise of significantly advancing our understanding of the physics and astrophysics of compact objects. Hence, this analysis targets a variety of GW signals including those produced by  non-axisymmetric deformations in newly-born NSs or magnetars formed in massive star core collapses or binary NS mergers \cite{GRBPlateau1,2017ApJ...843L...1L,2018PhRvD..98d3011S,2015ApJ...798...25D}, fallback accretion onto newly-born NSs or BHs \cite{1995ApJ...442..259L,2011ApJ...736..108P,2012ApJ...761...63P,2016ApJ...819..169V}, or instabilities and fragmentation in the accretion disks around BHs \cite{ADI}. Low-mass CBCs can also generate GW signals that are relatively long-lived ($\gtrsim 1$\,s) in the frequency band of ground-based GW detectors. Generally, these CBC signals are well-modeled, and therefore better searched for with matched filtering techniques \cite{2023MNRAS.524.5984L}.
However, here we also target low-mass (total mass $\leq5\,M_\odot$) CBCs with high eccentricity ($0.2 \le e \le 0.6$), as these do not fall within the parameter space covered by matched filter-based searches. Hereafter, we refer to these systems as eccentric compact binary coalescences (ECBCs).
Our results complement other dedicated waveform-independent searches for high mass (source total mass $M\geq 70\,M_\odot$) eccentric CBC signals with eccentricity $e \leq 0.3$ (e.g., \cite{2024ApJ...973..132A}).

This paper is organized as follows. In Section \ref{Sec:data} we describe our dataset. Sections \ref{Sec:method} and  \ref{Sec:waveforms} present our data analysis methods and the reference waveforms used to quantify our sensitivity, respectively. In Section \ref{Sec:results} and \ref{Sec:discussion} we report our results and conclude.

\section{Data}
\label{Sec:data}
The LIGO-Virgo-KAGRA fourth observing run (O4) started on 24 May 2023 at 15:00 UTC. The first part of the O4 observing run, O4a, ended on 16 January 2024 at 16:00\,UTC. 
During O4a, the LIGO Hanford (LHO) and Livingston (LLO) detectors
operated at improved sensitivity compared to their O3 run. A conventional measure of sensitivity is the binary neutron star (BNS) inspiral range, which quantifies the average distance at which a fiducial 1.4-1.4\,M$_{\odot}$ BNS could be detected with a signal-to-noise ratio (SNR) of 8 (see e.g., \citep{2021CQGra..38e5010C} and references therein). During O4a, the LHO and LLO detectors reached a BNS range of about 160 Mpc,
corresponding to approximately $ 30\%$ and $15\% $ improvements, respectively, compared to O3 \cite{2025PhRvD.111f2002C}. The Virgo detector did not join O4a.  

The search algorithms employed in this work require coincident data from at least two detectors. Because the BNS range for the KAGRA detector was substantially smaller than that of the LIGO detectors in O4a, we do not include KAGRA data in our search and we consider only data where both LIGO detectors are simultaneously available. A total of $126.6$\,days of coincident LHO-LLO data were collected during O4a. This corresponds to a duty cycle of about $53\%$ for joint observations collecting so-called ``ANALYSIS\_READY'' data. The last are data collected with the interferometers operating under observing conditions that are considered suitable for searching for GW signals. Removal of coincident data with significant data quality issues (by applying so-called ``category 1'' vetoes, as defined in \citep{2021CQGra..38m5014D}), left us with about $125$ days of coincident LHO-LLO data \cite{2025CQGra..42h5016S}.  Next, a small fraction (about $1-2$\%) of this coincident data is discarded because their duration is shorter than the time window used by each pipeline (see Section \ref{Sec:method} for details). Finally, as we describe in the next Section, strategies to reduce the impact of glitches (transient noise events that have a variety of origins), resulted in small ($\lesssim 1\%$) amount of data being removed from the analysis. 

During O4a, the calibration uncertainty in amplitude below 2\,kHz was less than 10\%, and it improved in the second half of O4a to be less than 2\% below 2\,kHz. Calibration uncertainty is not taken into account in the results presented in this paper as these calibration errors are much smaller than the astrophysical uncertainties that affect the class of signals we target here.

\section{Search Methods}
\label{Sec:method}

Given the large uncertainties in potential post-merger GW signal characteristics (Fig.\,\ref{fig:waveforms}; \cite{GRBPlateau1,2017ApJ...843L...1L,2018PhRvD..98d3011S,2015ApJ...798...25D,1995ApJ...442..259L,2011ApJ...736..108P,2012ApJ...761...63P,2016ApJ...819..169V,ADI}), hereafter we use two independent unmodeled search methods (and corresponding background estimations) that are based on different data-processing and clustering techniques: cWB \cite{cwb,XGBoost} and PySTAMPAS \citep{2021PhRvD.104j2005M}. In this Section, we discuss briefly their basic workings. More details are provided in Sections \ref{sec:cWB}-\ref{sec:pystampas}.

Unmodeled searches for long-duration GW transients typically look for patterns of excess power in some time-frequency representation of the data, and rely on the cross-correlation between the data streams of non co-located detectors to distinguish actual astrophysical GWs from background events generated by instrumental or environmental noise in the detectors \citep{STAMP,Lonetrack,Seedless1,Seedless2,Seedbased,PySTAMPAS,ALBUS,ALBUSglitches,CWB_longdur,2016PhRvD..93j4059C}. 

To estimate the distribution of background events, the data from one detector are time-shifted with respect to the other by an amount of time large enough to remove any coherent GW signal from the cross-correlated data. The process of time sliding the data is repeated multiple times for different values of the time shift to estimate the inverse false-alarm rate (iFAR) of potential candidate events accurately. 
\subsection{Coherent Wave Burst}
\label{sec:cWB}
cWB is an unmodeled transient search pipeline \citep{cwb}. It is based on a multi-resolution time-frequency wavelet transform known as Wilson-Daubechies-Meyer (WDM) \cite{Necula_2012}.  
Time-frequency pixels that contain excess energy (as estimated from the wavelet coefficients) across the detector network are selected and nearby pixels are clustered. In the pixel selection process, periods where a known physical factor is affecting the detector’s data quality are removed (so-called ``category 2'' vetoes, defined as in \citep{2021CQGra..38m5014D,2025CQGra..42h5016S}). 

After the clustering step, cWB performs an all-sky search for each cluster of pixels by utilizing a likelihood algorithm computed over an equal area pixelated grid of sky positions obtained using the Hierarchical Equal Area isoLatitude Pixelization (HEALPix\footnote{\url{http://healpix.sf.net}}) scheme \citep{Zonca2019,2005ApJ...622..759G}. Triggers are reconstructed from the sky position where the likelihood reaches a maximum. 

In the post-processing, cWB (version 6.4.5.0) uses a supervised boosted decision tree classifier, XGBoost \cite{XGBoost}, to re-weight the network coherent signal-to-noise ratio (SNR) of triggers.  The re-weighted values are used as ranking statistics \cite{PhysRevD.107.062002}. We note that this XGBoost re-weighting procedure entirely replaces the waveform duration cut (and all other thresholds in post-processing) used for cWB long-duration searches in O2 \citep{2018CQGra..35f5009A} and O3 \citep{2021PhRvD.104j2001A}. Eliminating the duration cut enables long-duration searches on a larger parameter space, while enhancing sensitivity. Consequently, some CBC triggers with high SNR identified by cWB during the search retain high ranking statistics after the re-weighting process. Hence, similar to the approach employed for short-duration signals \cite{O4ashortburst}, CBC signals are excised a posteriori from the analysis (see Section \ref{Sec:results} and Figure\,\ref{fig:counts_vs_IFAR} for more details).

In the O4a long-duration analysis, the data are divided into four chunks of approximately equal duration across the approximately $122$ days of observing time analyzed. The chunks are further split into 1200\,s-long data segments, which are then transformed with the WDM wavelet into seven time-frequency resolutions (frequency bins ranging from 0.5\,Hz to 16\,Hz; time bins ranging from 1/64\,s to 1\,s) covering the frequency range 16--2048 Hz.
To estimate the background distribution of noise triggers, time slides are performed by sliding the data between detectors 600 times in 2-second steps within each 1200s-long data segment. A 2-second time step is longer than the longest time bin utilized for the cWB search, and sufficient to break any coherence in the time-frequency pixels. This process is further extended by performing shifts across different data segments of each chunk, resulting in a total of 2400 independent time shifts per chunk. This yields a background of about $772$\,years in total for the four chunks (half of which is used for training, as described below). 

The XGBoost models for each chunk are trained independently, with dedicated training sets. These training sets are created by randomly selecting 50\% of the background data from the corresponding chunk, and injecting a series of white noise burst signals to cover the analysis frequency range. The produced XGBoost model is then applied to the remaining 50\% background to produce the background statistical distribution. To evaluate the efficiency of the XGBoost model, a set of simulations with a selection of waveform models, as described in Section \ref{Sec:waveforms}, is analyzed with the XGBoost models. 

We note that in O4a, excessive noise fluctuations around the known 60\,Hz and 180\,Hz power lines and harmonics are observed in the background. Thus, in the post-production, noise fluctuations at these frequencies are excised (from both the background and the foreground). 

\subsection{PySTAMPAS}
\label{sec:pystampas}
PySTAMPAS \citep{2021PhRvD.104j2005M} is an enhanced version of the Stochastic Transient Analysis Multi-detector Pipeline (STAMP) \citep{2011PhRvD..83h3004T} that was used in previous analyses \citep{2019PhRvD..99j4033A, 2021PhRvD.104j2001A}. It is designed to perform unmodeled all-sky searches for long-duration GW transients at a reduced computational cost compared to STAMP \citep{2021PhRvD.104j2005M}. The data are split into 512\,s-long windows which overlap by 50\%. For each window and each detector, spectrograms of the auto-power SNR are built using the Short-Time Fourier Transform (STFT) over short segments with duration 0.5\,s, 1\,s, 2\,s, and 4\,s, that are Hann-windowed and overlap by 50\%. The four spectrograms are then combined into a single multi-resolution spectrogram that covers the frequency band 22--2000\,Hz. For each frequency bin, the power spectral density (PSD) is estimated by taking the median of the squared modulus of the STFT over the window duration. A seed-based clustering algorithm is applied on each multi-resolution spectrogram to extract clusters of pixels that form candidate GW triggers. Pixels forming each cluster are then cross-correlated with the spectrogram from the other detector and a coherent detection statistic is computed for the whole trigger \citep{2021PhRvD.104j2001A}. 

To estimate the background distribution of noise triggers, the spectrogram from the other detector is time-shifted by an amount of time greater than 1000\,s. This is a conservative choice that allows us to break any coherence in time over the duration of the longest signals targeted by this search, regardless of signal frequency content. This operation is repeated $320$ times with different values of the time shift, allowing to simulate $108$ years of background.
To deal with non-Gaussian noise artifacts, frequency bins that correspond to known spectral lines in the detectors and that generate an excess of noise triggers, are masked. In total, $9$\% of the total frequency band is masked in this analysis. 
As was done in previous searches with STAMP \citep{2021PhRvD.104j2001A}, triggers for which the maximal fraction of SNR in a single time bin exceeds 0.3, or for which the ratio of auto-power SNR between the two detectors exceeds $5$, or that have a duration lower than $10$\,s, are vetoed in post-processing. Because of the high rate of noise fluctuations at low frequency, triggers whose central frequency is below $50$\,Hz are also dismissed. Finally, spectrograms for which the fraction of pixels above the threshold considered for clustering is above $0.5\%$ are removed from the analysis. These correspond to unusually noisy stretches of data that the clustering algorithm is unable to handle. About $0.03\%$ of the O4 data considered for this search (see Section \ref{Sec:data}) are removed by this cut.

\begin{figure}
\label{Fig:waveforms}
    \centering
    \includegraphics[width=\linewidth]{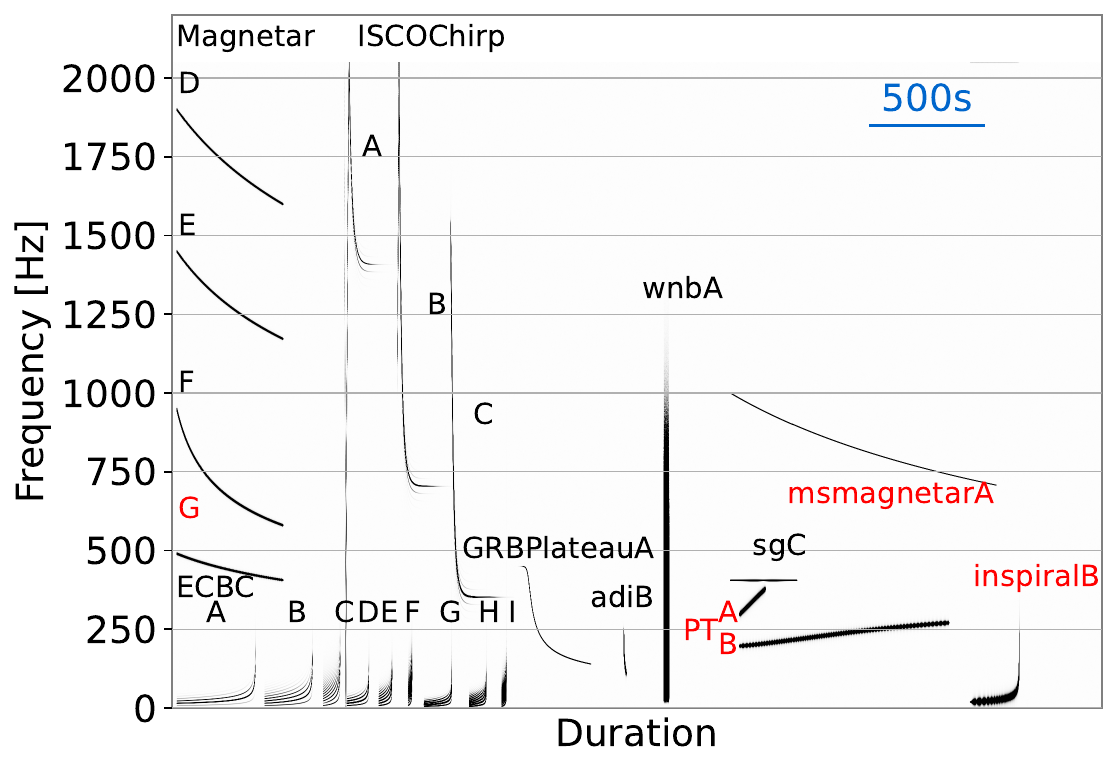}
    \caption{Time-frequency representation of the waveforms used to test the sensitivity of this search. The x-axis represents a linear time axis with the various signal types off-set horizontally for clarity. Waveforms that are new in this analysis (compared to  O2 \citep{2017ApJ...851L..16A,2019PhRvD..99j4033A} and O3 \citep{2021PhRvD.104j2001A} long-duration burst searches) are marked in red.
\label{fig:waveforms}}
\end{figure}

\section{Waveform models}
\label{Sec:waveforms}
The analysis presented here is an unmodeled search for GW signals, and as such it does not rely on the use of template waveforms to make detection statements. In the absence of a detection, to put our results in context, we quantify the sensitivity of our search by setting upper limits on the GW strain amplitude of
a set of simulated waveforms added coherently into detector data. We note that because of the limited distance reach of our analysis, we do not consider redshift effects on the model waveforms.

The waveform models used in this analysis include those from a similar search performed on O3 data \citep{2021PhRvD.104j2001A}. These waveforms are representative of some compelling astrophysical scenarios, such as post-merger magnetars (Magnetar) with ellipticity in the range $0.005 - 0.08$,\citep{2015ApJ...798...25D},  accretion disk instabilities (ADI) \citep{ADI}, newly formed magnetars powering gamma-ray burst plateaus (GRBplateau)  \citep{1995ApJ...442..259L,GRBPlateau1,GRBPlateau2}, inspiral-merger-ringdown ECBC waveforms   \citep{2018PhRvD..97b4031H} with eccentricity between $0.2$ and $0.6$ and total mass between $2.8$ and $10\,M_\odot$, and broadband chirps from innermost stable circular orbit waves around rotating BHs with mass between $5$ and $20$ $M_\odot$ (ISCOchirp) \citep{2016ApJ...819..169V}. 

In addition to the above, we include in our analysis GW signals modeled for binary NS mergers (inspiralB), millisecond magnetars (msmagnetar-A), and fallback accretion onto NSs (PT-A; PT-B) with their time frequency representation shown in Figure\,\ref{fig:waveforms}. The inspiral model employs the IMRPhenomPv2NRTidalv2 approximant for equal-mass binaries (1.4 \(M_\odot\)) \citep{2019PhRvD.100d4003D}. The millisecond magnetar model is an analytical model derived from the dynamics of spinning down nascent NSs proposed by \citet{2018PhRvD..98d3011S} and \citet{2017ApJ...843L...1L} with frequency evolution modeled with an arbitrary but fixed braking index. The PT models (A and B) were proposed by \citet{2012ApJ...761...63P} for stars of intermediate mass that end their lives forming NSs which eventually collapse to BHs via fallback accretion. When the incoming material has sufficient angular momentum to form a disk, the accretion spins up the NS sufficiently to produce non-axisymmetric instabilities and gravitational radiation with frequencies in the range $ 700$--2400\,Hz for about $30$--3000\,s until collapse to a BH occurs.

Finally, to fill out the parameter space, we perform sensitivity estimates also for ``ad-hoc'' waveforms: band-limited white noise burst (WNB) and sine-Gaussian bursts (SG).

\section{Results}
\label{Sec:results}

Similarly to what was done in the O2 and O3 long-duration GW searches \citep{2019PhRvD..99j4033A,2021PhRvD.104j2001A}, for both the PySTAMPAS and cWB analyses we set a detection threshold corresponding to an  iFAR higher than 50\,years. 
\begin{figure}
    \centering
    \includegraphics[width=\linewidth]{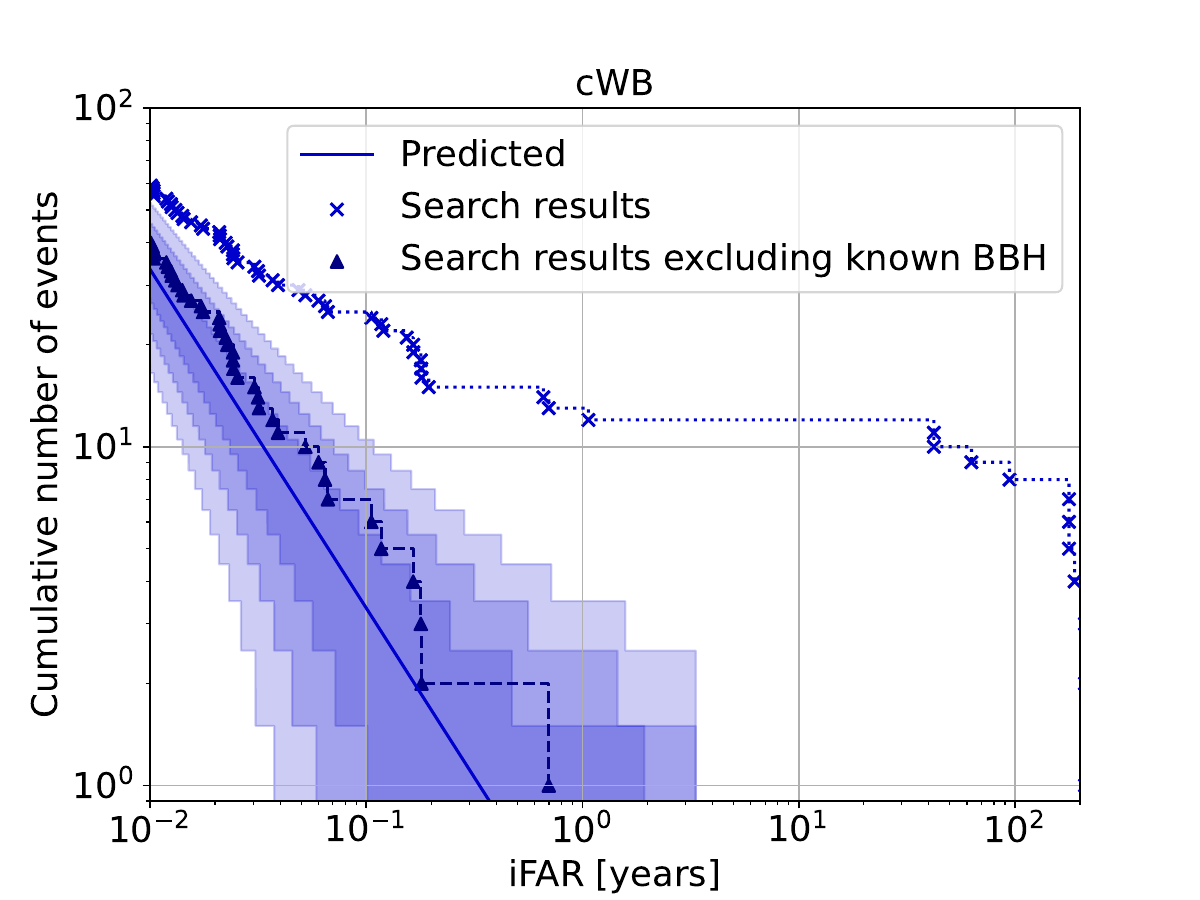}
    \includegraphics[width=\linewidth]{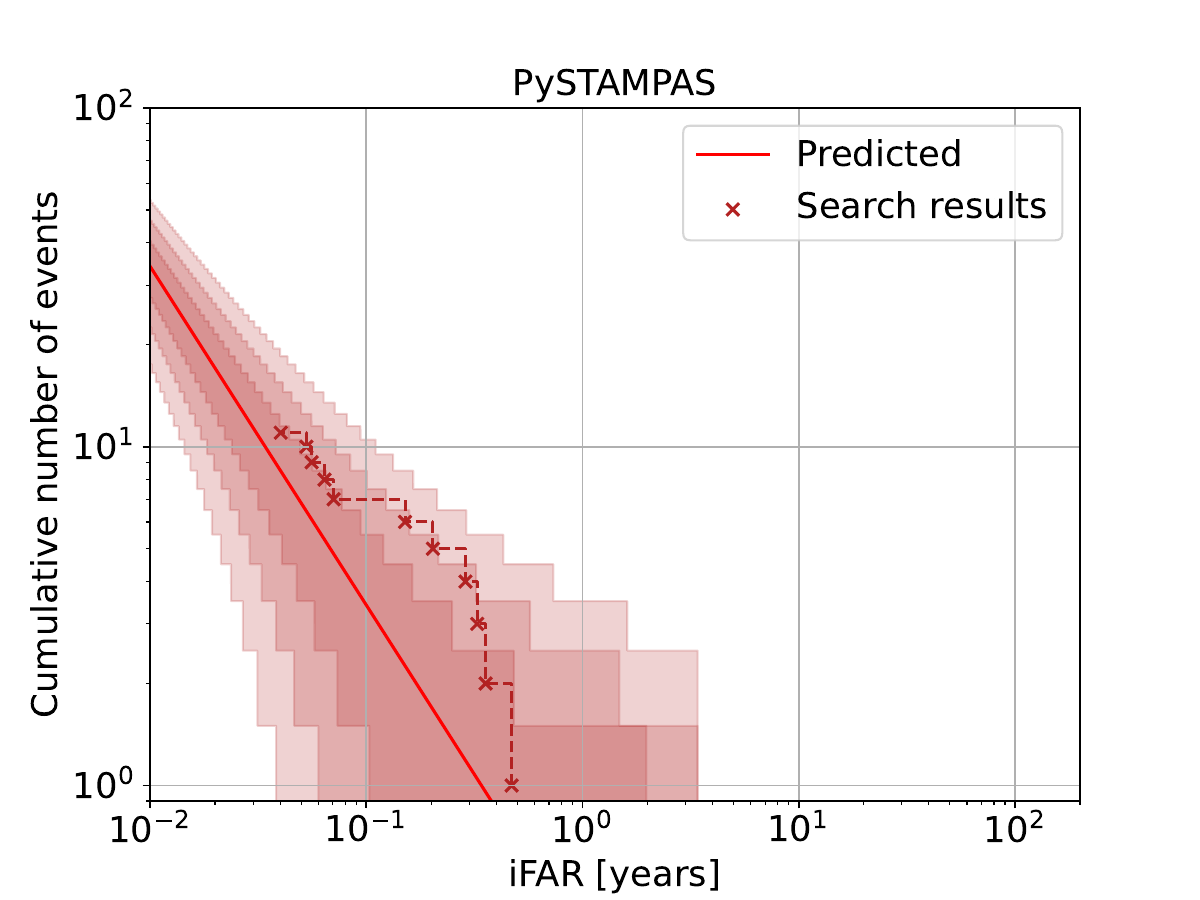}
    \caption{Cumulative number of events as a function of the iFAR found by cWB (top) and PySTAMPAS (bottom). Cross markers represent all events found in the search, while triangle ones exclude known BBH events found in low latency. Assuming a Poisson distribution for noise events, the expected value for the background (i.e., $T$/iFAR, where $T$ is the observing time) is shown by the solid line, while the shaded regions represents the $1-2-3 \, \sigma$ uncertainties. We note that the $\approx 2\sigma$ excess in the PySTAMPAS distribution over several iFAR bins is dominated by the contribution of the three loudest background events with FAP approximately $0.52$, $0.62$, and $0.65$, respectively.}
    \label{fig:counts_vs_IFAR}
\end{figure}

In the cWB analysis, we find 25 CBC triggers (11 of which with iFAR above the 50\,years detection threshold) that were confidently detected by low-latency searches, and subsequently reported as public alerts \citep{2023GCN.33917....1L,2023GCN.33914....1L,2023GCN.34075....1L,2023GCN.34086....1L,2023GCN.34113....1L,2023GCN.34161....1L,2023GCN.34380....1L,2023GCN.34411....1L,2023GCN.34504....1L,2023GCN.34692....1L,2023GCN.34741....1L,2023GCN.34775....1L,2023GCN.34739....1L,2023GCN.34756....1L,2023GCN.34760....1L,2023GCN.34799....1L,2023GCN.34895....1L,2023GCN.34927....1L,2023GCN.34967....1L,2023GCN.35168....1L,2023GCN.35297....1L,2023GCN.35298....1L,2023GCN.35420....1L,2023GCN.35428....1B,2024GCN.35493....1L}. After excluding triggers associated with these known, quasi-circular CBC events, the cWB search remains sensitive to ECBC signals. As shown in Figure~\ref{fig:counts_vs_IFAR}, the resulting distribution of cWB triggers is consistent with the background within $2 \sigma$. The most significant trigger has a SNR of 11, centered at 54\,Hz, with an iFAR of 0.70 years and a false alarm probability (FAP) of 0.38. This trigger happened during a period of elevated glitch rate. This, combined with the estimated FAP of $0.38$, suggests that the trigger is likely due to noise.

The distribution of triggers found by PySTAMPAS is also consistent with the background within $2 \sigma$, as shown in the bottom panel of Figure~\ref{fig:counts_vs_IFAR}. The most significant trigger has an iFAR of $0.49$ years (FAP $\approx 0.52$), and is consistent with a noise fluctuation in LLO between $242$\,Hz and $250$\,Hz. PySTAMPAS did not recover any of the CBC triggers identified by cWB because these had durations shorter than $10$\,s and are therefore removed in post-processing.

In the absence of a significant GW candidate in either analysis, we derive sensitivity estimates on the GW amplitude of the various waveform models described in Section \ref{Sec:waveforms}. To this end, simulated waveforms are injected coherently into the detectors' data at different amplitudes, and the detection efficiency (fraction of the total number of injections that are recovered) is measured as a function of waveform amplitude. The spanned range of amplitudes is chosen so that the detection efficiency is sampled across the whole range of its possible values---from $0\%$ to $100\%$. For each injected signal of a given amplitude, the starting time, sky position (right ascension and cosine of the declination), polarization angle, and cosine of the inclination angle are randomly drawn following a uniform distribution. We note that for estimating the detection efficiency, a signal is considered to be recovered by a given pipeline if it produces a trigger within the time and frequency boundaries of the injected waveform, with an iFAR higher than $50$ years. 

\begin{figure*}
    \centering
    \includegraphics[width=\linewidth]{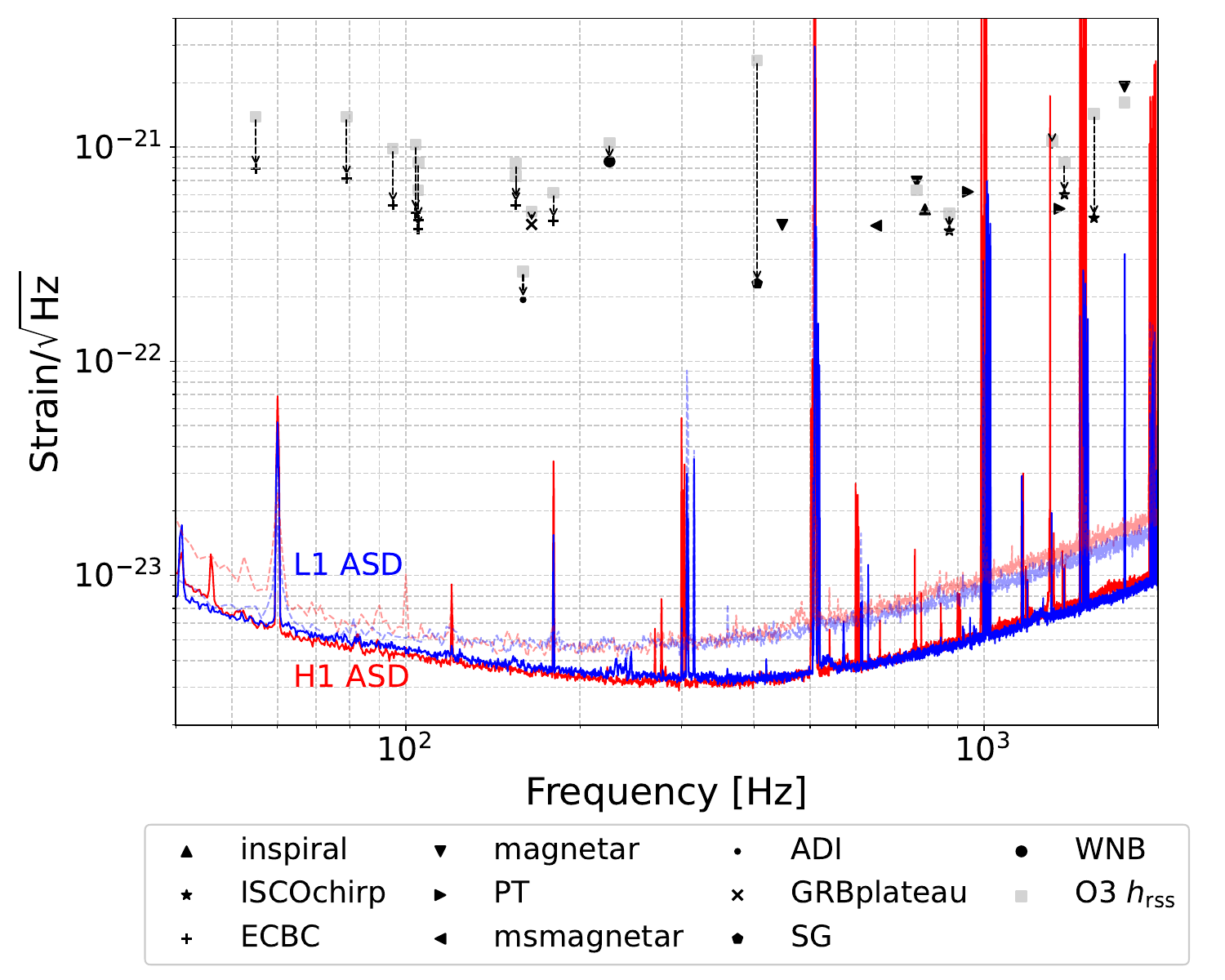}
    \caption{Root-sum-square amplitude at $50\%$ detection efficiency as a function of the central frequency of each tested waveform with iFAR\,$>50$\,years (black markers). The O3 results \citep{2021PhRvD.104j2001A} are represented by light-gray square markers, showing the improvement in sensitivity (vertical arrows). For each waveform, the most constraining result from either of the two pipelines is shown. For reference, the mean amplitude spectral densities of the Livingston (blue) and Hanford (red) detectors during O4a are plotted, along with those from the O3b (faded dashed curves). 
    \label{fig:sensitivity}}
\end{figure*}

The results are presented in Fig.\,\ref{fig:sensitivity} as root-sum-square amplitudes ($h_{\mathrm{rss}}$) at 50\% detection efficiency, where: \begin{equation}
h_{\rm rss} = \sqrt{\int_{-\infty}^{\infty} (h_{+}^2(t)+h_{\times}^2(t)) dt}.
\end{equation}
In the above equation, $h_+$ and $h_\times$ are the GW amplitudes for the $+$ and $\times$ polarizations. For each waveform, the lowest value from either of the two pipelines is reported. We discuss these results in Section \ref{Sec:discussion}.

We also derive upper limits on the rate of eccentric CBC events (ECBCs), updating previous results from O2 \cite{2019PhRvD..99j4033A} and O3 \cite{2021PhRvD.104j2001A}. 
In the absence of a detection, assuming that these events are uniformly distributed in the observed volume and follow a Poisson distribution, the $90\%$ confidence upper limit on their rates is given by \cite{2004CQGra..21S1775B}
\begin{equation}
\label{eqn:r90}
\mathcal{R}_{90 \%} = \frac{2.3}{4 \pi T \int_{0}^{r_{\mathrm{max}}}{dr \,r^2 \, \epsilon(r)}},
\end{equation}
where $\epsilon(r)$ is the detection efficiency as a function of distance $r$, $r_\textrm{max}$ is the maximal detectable distance, and $T$ is the observing time \cite{2012PhRvD..85l2007A}.
In Table~\ref{table:r90} we report our best results, derived using the cWB pipeline (which has the best sensitivity for this family of waveforms). We discuss these results in Section \ref{Sec:discussion}.

\begin{table}
\caption{Rate upper limits per unit volume at 90\% confidence level on eccentric CBC with various component masses masses and eccentricity $e$, computed with Eq.
\ref{eqn:r90}. In the last column, we show updated results from the O3 run, which we have recomputed as discussed in Section \ref{Sec:discussion}. We find that uncertainties on these values are dominated by systematic errors associated to the method used to fit the efficiency curves, estimated to be of order $15\%$ (see text for discussion).}
\label{table:r90}
\begin{tabular}{cccccc}
\hline
\multirow{2}{*}{Waveform} & \multirow{2}{*}{ $M_1 [M_{\odot}]$} & \multirow{2}{*}{$M_2 [M_{\odot}]$} & \multirow{2}{*}{$e$ } & \multicolumn{2}{c}{$\mathcal{R}_{90 \%}$ [$\rm{Gpc}^{-3} \rm{yr}^{-1}]$} \\
                  &                   &                   &                   &   O4a        &    O3      \\
            \hline
 ECBC\_A  & $1.4$ & $1.4$ &  $0.2$ &$9.1 \times 10^{3}$ & $2.0 \times 10^{4}$\\
 ECBC\_B  & $1.4$ & $1.4$ &  $0.4$ &$1.1 \times 10^{4}$ & $2.4 \times 10^{4}$ \\
 ECBC\_C  & $1.4$ & $1.4$ &  $0.6$ &$1.8 \times 10^{4}$ & $4.4 \times 10^{4}$\\
 ECBC\_D  & $3.0$ & $3.0$ &  $0.2$ &$1.3 \times 10^{3}$ & $4.6 \times 10^{3}$\\
 ECBC\_E  & $3.0$ & $3.0$ &  $0.4$ &$1.4 \times 10^{3}$ & $5.5 \times 10^{3}$\\
 ECBC\_F  & $3.0$ & $3.0$ &  $0.6$ &$3.7 \times 10^{3}$ & $8.9 \times 10^{3}$\\
 ECBC\_G  & $5.0$ & $5.0$ &  $0.2$  &$4.2 \times 10^{2}$ & $3.0 \times 10^{3}$\\
 ECBC\_H  & $5.0$ & $5.0$ &  $0.4$ &$5.2\times 10^{2}$ & $4.0 \times 10^{3}$\\
 ECBC\_I  & $5.0$ & $5.0$ &  $0.6$ &$8.0 \times 10^{2}$ & $4.7 \times 10^{3}$\\
\hline 
\end{tabular}
\end{table}

\section{Discussion}
\label{Sec:discussion}

The $h_{\rm rss}$ at 50\% efficiency of this O4a search has decreased by 30\% (on average across the detector bandwidth) compared to O3 (Figure \ref{fig:sensitivity}).
This improvement is primarily driven by the sensitivity increase of the LIGO detectors between O3 and O4a. However, the pipelines used for this search are also different from the ones used in O3. This improves the O4a sensitivity (compared to O3) for most waveforms. Specifically, the ECBCs are better recovered than in O3 (up to a factor about 2 lower in $h_{\rm rss}$ at 50\% efficiency), because the new version of cWB, enhanced with XGBoost, better discriminates these signals from noise transients \citep{XGBoost}. 
The sensitivity to monochromatic SGs has also improved by a factor of about 10 thanks to a more robust PSD estimation in PySTAMPAS (which is now based on the median PSD over a time window longer than twice the duration of the signal). The sensitivity to magnetar waveforms has not improved compared to O3, as these signals are best recovered using the seedless clustering strategy used in O3 \cite{Seedless1}. The seedless clustering strategy is being developed in the PySTAMPAS package at the time of writing. Hence, this strategy has not been used for the O4a results presented here.

To put the above results in the context of previous analyses, we note that the O3 long-duration search broadly constrained GW signals with energies of order $10^{-2}\,M_{\odot}$ (comparable to the maximum rotational energy of a $1.4\,M_{\odot}$ NS \citep{2017MNRAS.472.3058B}), and morphologies similar to the ones considered here, to distances $\gtrsim 1-10$\,Mpc (see Figure  2 in \citep{2021PhRvD.104j2001A}). 
For a GW signal with an energy $E_\textrm{GW}$ and central frequency $f_0$, the horizon distance $d$ corresponding to a given $h_\textrm{rss}$ is $d \simeq \sqrt{\frac{5 G E_\textrm{GW}}{2 \pi^2 c^3 f_0^2 h_{\textrm{rss}}^2}}$ \cite{2013arXiv1304.0210S}.
The overall improvement of $\approx 30\%$ in $h_{\rm rss}$ at 50\% efficiency achieved via this analysis pushes the above lower limits on the distances to values about $30\%$ larger. 

We also derive rates upper limits on ECBC events (Table \ref{table:r90}). The upper limits range from $420\,\rm{Gpc}^{-3} \rm{yr}^{-1}$ to $18,000\, \rm{Gpc}^{-3} \rm{yr}^{-1}$ depending on the mass and eccentricity of the system. 
These ECBC rate constraints are compatible with expectations that, overall, only a small
fraction of CBC systems have a significant eccentricity~\cite{PhysRev.136.B1224}, given current estimates of CBC rates ($10-1,700 \: \rm{Gpc}^{-3} \rm{yr}^{-1}$ for NS-NS systems, $7.8-140 \: \rm{Gpc}^{-3} \rm{yr}^{-1}$ for NS-BH, and $17.9-44\: \rm{Gpc}^{-3} \rm{yr}^{-1}$ for BH-BH \cite{2023PhRvX..13a1048A}).

\begin{figure}
    \centering
    \includegraphics[width=\linewidth]{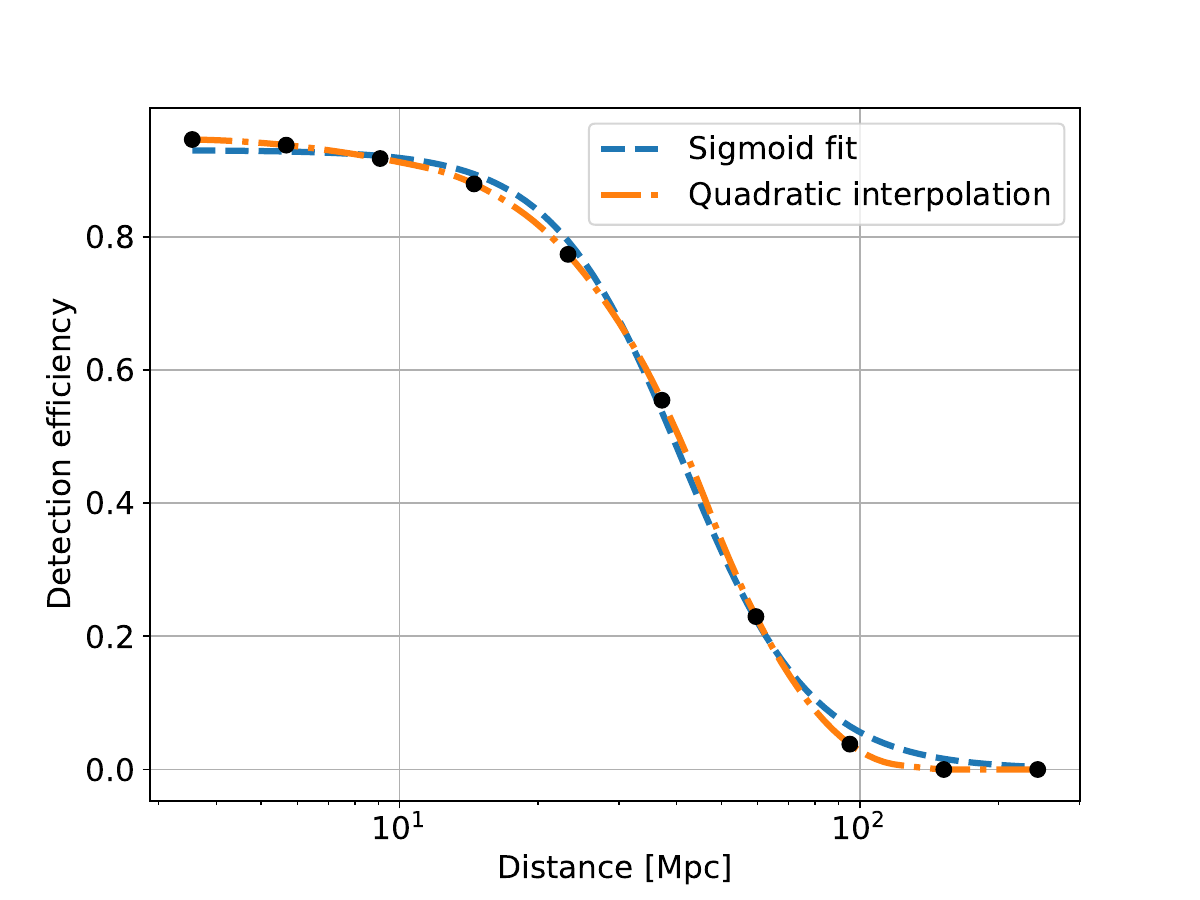}
    \caption{Detection efficiency as a function of the distance recovered by cWB for the ECBC-A waveform model on O4a data. The black dots represent data points, and the dashed and dashed dotted curves represent a fit by a sigmoid function and a quadratic interpolation respectively. This illustrates how a sigmoid fit overestimates the detection efficiency at large distances, leading to a detection volume overestimated by a factor $\sim 2$ in that case.}
    \label{fig:efficiency_curve}

\end{figure}

The above rate upper limits ECBCs require 
a careful sampling and modeling of the tail of the detection efficiency curve at large distances. For this analysis (O4a column in Table\,\ref{table:r90}), 
we revisit and improve the efficiency estimation procedure that was employed in O3. First, we use a quadratic interpolation of the sampled efficiency values instead of the sigmoid fit used in O3 \cite{2021PhRvD.104j2001A}. To assess the robustness of the quadratic interpolation, we compare our results with a log-normal distribution fit and a linear interpolation. Both these methods give results consistent with the quadratic interpolation within about 15\% of the estimated rate upper limit.
On the other hand, a sigmoid fit  systematically overestimates the detection volume (the integral in the denominator of Eq.\,\ref{eqn:r90}), resulting in an underestimate of the 90\% rate upper limits by a factor of about 2--3 across different waveforms and detection pipelines. An example of this is presented in Figure \ref{fig:efficiency_curve}.
As a second improvement to our detection efficiency estimates,  we limit the $r_\mathrm{max}$ in Eq. \ref{eqn:r90} to the highest sampled distance so as to ensure that the 90\% rate upper-limits are not affected by errors that may be introduced when extrapolating fits or interpolations to the efficiency curves at distances beyond the range that was actually sampled.  
In the O3 long duration search \cite{2021PhRvD.104j2001A},  the lack of the above constraint 
resulted in an underestimation of the rate upper limits by factor of about $5$ (on average across ECBC waveforms) for the cWB pipeline. This issue did not affect the O3 
STAMP-AS Zebragard results \citep{2021PhRvD.104j2001A}. 

Given the above noted improvements in the detection volume estimates, in the rightmost column of Table \ref{table:r90} we compare our O4a results with the rate upper limits that we derive applying the same efficiency curve estimation of our O4a analysis to the (previously unpublished) O3 STAMP-AS Zebragard results for ECBCs. The O4a upper limits improve on O3 results by factors of about $2-7$, consistently with the improvement in sensitivity shown in Fig.\,\ref{fig:sensitivity} and the fact that the observing time of O4a is about half of O3. Indeed, for this waveform family, the improved O4a sensitivity implies a distance reach about $1.4-2.4$ times larger than in O3, hence a detection volume about $2.7-14$ times larger, and therefore rate upper limits $1.4-7$ times smaller. 

In conclusion, with this O4a analysis we have made substantial progress in our ability to search for long-duration GW signals,
thanks to the combination of the improved O4a sensitivity and enhancements in pipeline detection efficiencies. With further progress in GW detectors' sensitivity \citep{2018LRR....21....3A} and longer data taking runs, long-duration unmodeled searches have the potential to probe several interesting astrophysical scenarios involving NSs and BHs \cite{GRBPlateau1,2017ApJ...843L...1L,2018PhRvD..98d3011S,2015ApJ...798...25D,1995ApJ...442..259L,2011ApJ...736..108P,2012ApJ...761...63P,2016ApJ...819..169V,ADI}. We note that the pipeline improvements described in this search can also benefit triggered searches for long-duration post-merger signals  \citep{2017ApJ...851L..16A,2019ApJ...880L..15M}, addressing a key open question left open by the multi-messenger detection of GW170817, namely, what is the nature of the post-merger remnant.

\acknowledgements 
%
This material is based upon work supported by NSF's LIGO Laboratory, which is a
major facility fully funded by the National Science Foundation.
The authors also gratefully acknowledge the support of
the Science and Technology Facilities Council (STFC) of the
United Kingdom, the Max-Planck-Society (MPS), and the State of
Niedersachsen/Germany for support of the construction of Advanced LIGO 
and construction and operation of the GEO\,600 detector. 
Additional support for Advanced LIGO was provided by the Australian Research Council.
The authors gratefully acknowledge the Italian Istituto Nazionale di Fisica Nucleare (INFN),  
the French Centre National de la Recherche Scientifique (CNRS) and
the Netherlands Organization for Scientific Research (NWO)
for the construction and operation of the Virgo detector
and the creation and support  of the EGO consortium. 
The authors also gratefully acknowledge research support from these agencies as well as by 
the Council of Scientific and Industrial Research of India, 
the Department of Science and Technology, India,
the Science \& Engineering Research Board (SERB), India,
the Ministry of Human Resource Development, India,
the Spanish Agencia Estatal de Investigaci\'on (AEI),
the Spanish Ministerio de Ciencia, Innovaci\'on y Universidades,
the European Union NextGenerationEU/PRTR (PRTR-C17.I1),
the ICSC - CentroNazionale di Ricerca in High Performance Computing, Big Data
and Quantum Computing, funded by the European Union NextGenerationEU,
the Comunitat Auton\`oma de les Illes Balears through the Conselleria d'Educaci\'o i Universitats,
the Conselleria d'Innovaci\'o, Universitats, Ci\`encia i Societat Digital de la Generalitat Valenciana and
the CERCA Programme Generalitat de Catalunya, Spain,
the Polish National Agency for Academic Exchange,
the National Science Centre of Poland and the European Union - European Regional
Development Fund;
the Foundation for Polish Science (FNP),
the Polish Ministry of Science and Higher Education,
the Swiss National Science Foundation (SNSF),
the Russian Science Foundation,
the European Commission,
the European Social Funds (ESF),
the European Regional Development Funds (ERDF),
the Royal Society, 
the Scottish Funding Council, 
the Scottish Universities Physics Alliance, 
the Hungarian Scientific Research Fund (OTKA),
the French Lyon Institute of Origins (LIO),
the Belgian Fonds de la Recherche Scientifique (FRS-FNRS), 
Actions de Recherche Concert\'ees (ARC) and
Fonds Wetenschappelijk Onderzoek - Vlaanderen (FWO), Belgium,
the Paris \^{I}le-de-France Region, 
the National Research, Development and Innovation Office of Hungary (NKFIH), 
the National Research Foundation of Korea,
the Natural Sciences and Engineering Research Council of Canada (NSERC),
the Canadian Foundation for Innovation (CFI),
the Brazilian Ministry of Science, Technology, and Innovations,
the International Center for Theoretical Physics South American Institute for Fundamental Research (ICTP-SAIFR), 
the Research Grants Council of Hong Kong,
the National Natural Science Foundation of China (NSFC),
the Israel Science Foundation (ISF),
the US-Israel Binational Science Fund (BSF),
the Leverhulme Trust, 
the Research Corporation,
the National Science and Technology Council (NSTC), Taiwan,
the United States Department of Energy,
and
the Kavli Foundation.
The authors gratefully acknowledge the support of the NSF, STFC, INFN and CNRS for provision of computational resources.

This work was supported by MEXT,
the JSPS Leading-edge Research Infrastructure Program,
JSPS Grant-in-Aid for Specially Promoted Research 26000005,
JSPS Grant-in-Aid for Scientific Research on Innovative Areas 2402: 24103006,
24103005, and 2905: JP17H06358, JP17H06361 and JP17H06364,
JSPS Core-to-Core Program A.\ Advanced Research Networks,
JSPS Grants-in-Aid for Scientific Research (S) 17H06133 and 20H05639,
JSPS Grant-in-Aid for Transformative Research Areas (A) 20A203: JP20H05854,
the joint research program of the Institute for Cosmic Ray Research,
University of Tokyo,
the National Research Foundation (NRF),
the Computing Infrastructure Project of the Global Science experimental Data hub
Center (GSDC) at KISTI,
the Korea Astronomy and Space Science Institute (KASI),
the Ministry of Science and ICT (MSIT) in Korea,
Academia Sinica (AS),
the AS Grid Center (ASGC) and the National Science and Technology Council (NSTC)
in Taiwan under grants including the Science Vanguard Research Program,
the Advanced Technology Center (ATC) of NAOJ,
and the Mechanical Engineering Center of KEK.

Additional acknowledgements for support of individual authors may be found in the following document: \\
\texttt{https://dcc.ligo.org/LIGO-M2300033/public}.
For the purpose of open access, the authors have applied a Creative Commons Attribution (CC BY)
license to any Author Accepted Manuscript version arising.
We request that citations to this article use 'A. G. Abac {\it et al.} (LIGO-Virgo-KAGRA Collaboration), ...' or similar phrasing, depending on journal convention.




\bibliographystyle{apsrev4-2}
\bibliography{references}

\end{document}